\documentclass[a4paper]{article}

\pdfoutput=1

\usepackage[utf8]{inputenc}
\usepackage[T1]{fontenc}
\usepackage{lmodern}
\usepackage{graphicx}
\usepackage{fancyvrb}
\usepackage{multirow}
\usepackage{hyperref}
\usepackage{fullpage}
\begin{document}
\pagestyle{empty}
\begin{center}
\begin{figure}%
\centering
\includegraphics[width=0.5\columnwidth]{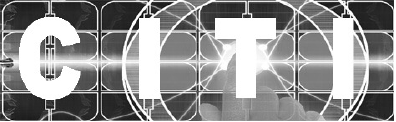}%
\end{figure}
{\LARGE \textsc{Scientific Communication}} \\
\vspace{0.5cm}
{Centre of Innovation in Telecommunications and Integration of services} \\
\vspace{3cm}
{\Large \textbf{JooFlux: Hijacking Java 7 InvokeDynamic \\ To Support Live Code Modifications}} \\
\vspace{10pt}
{\large \textit{Julien Ponge \& Frédéric Le Mouël}} \\
\vspace{10pt}
{\large October 2012} \\
\vspace{20pt}
\begin{minipage}{0.8\columnwidth}
\sffamily
\small
Changing functional and non-functional software implementation at runtime is useful
and even sometimes critical both in development and production environments.  JooFlux is a JVM agent 
that allows both the dynamic replacement of method implementations and the application of aspect 
advices. It works by doing bytecode transformation to take advantage of the new \texttt{invokedynamic} 
instruction added in Java SE 7 to help implementing dynamic languages for the JVM. JooFlux can be managed 
using a JMX agent so as to operate dynamic modifications at runtime, without resorting to a dedicated
domain-specific language. We compared JooFlux with existing AOP platforms and dynamic languages.
Results demonstrate that JooFlux performances are close to the Java ones --- with most of the time a
marginal overhead, and sometimes a gain --- where AOP platforms and dynamic languages present
significant overheads. This paves the way for interesting future evolutions and applications of
JooFlux.
\end{minipage}
\end{center}
\vfill
\begin{minipage}[b]{0.3\columnwidth}
\includegraphics[width=\columnwidth]{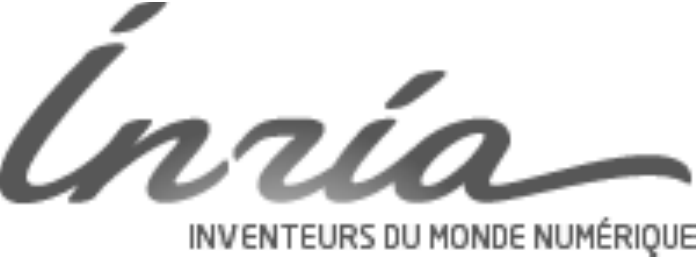}%
\end{minipage}
\hfill
\begin{minipage}{0.3\columnwidth}
\small
\centering
\textbf{
University of Lyon \\
INSA Lyon \\
INRIA
}
\end{minipage}
\hfill
\begin{minipage}[b]{0.25\columnwidth}
\includegraphics[width=\columnwidth]{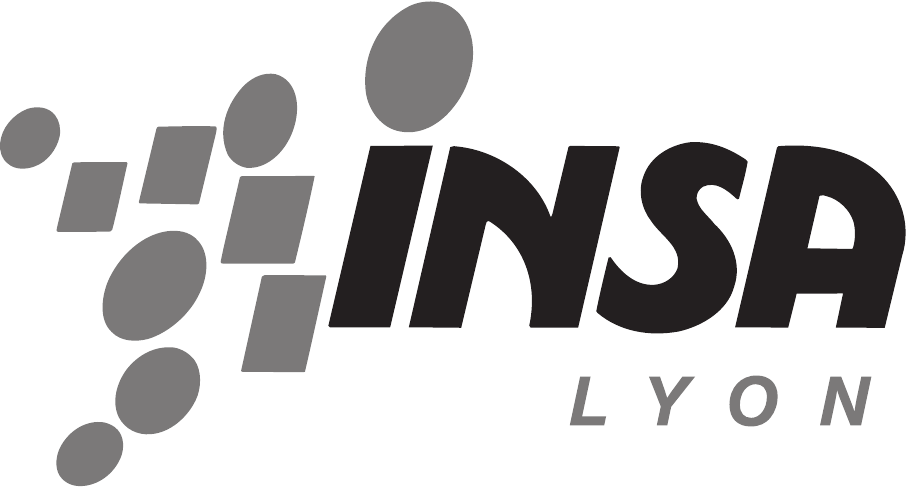}%
\end{minipage}
\newpage

\title{JooFlux: Hijacking Java 7 InvokeDynamic \\ To Support Live Code Modifications}

\author{Julien Ponge and Frédéric Le Mouël\\[10pt]
University of Lyon\\
INSA-Lyon, CITI-INRIA F-69621, Villeurbanne, France\\
\texttt{julien.ponge@insa-lyon.fr} and \texttt{frederic.le-mouel@insa-lyon.fr}}

\date{October 2, 2012}

\maketitle

\begin{abstract} Changing functional and non-functional software implementation at runtime is useful
and even sometimes critical both in development and production environments.  JooFlux is a JVM agent 
that allows both the dynamic replacement of method implementations and the application of aspect 
advices. It works by doing bytecode transformation to take advantage of the new \texttt{invokedynamic} 
instruction added in Java SE 7 to help implementing dynamic languages for the JVM. JooFlux can be managed 
using a JMX agent so as to operate dynamic modifications at runtime, without resorting to a dedicated
domain-specific language. We compared JooFlux with existing AOP platforms and dynamic languages.
Results demonstrate that JooFlux performances are close to the Java ones --- with most of the time a
marginal overhead, and sometimes a gain --- where AOP platforms and dynamic languages present
significant overheads. This paves the way for interesting future evolutions and applications of
JooFlux.
\end{abstract}

\newpage

\section{Introduction}

Most applications are --- to some degree --- relatively static. This is especially true for software
developed using statically typed languages. Most dynamically-typed languages offer the ability to
replace methods at runtime, also coined as \emph{monkey patching}. Even so, this is rarely possible
to perform from outside the program: such changes can be performed by the program, not by some out
of process management tool. Very few languages make it easy to do so from the outside, even purely
functional languages with \emph{read-eval-print loops}.

Dynamically changing code implementation at runtime has many virtues
though~\cite{Frieder1991,Hicks2005}. In development phases, a significant share of time is lost due
to the need to restart applications to see the effect of source code changes. In production,
applying an important security or bug fix requires the same restart procedure, this time impacting
the deployed application availability and causing issues to end-user. A full restart can take
several minutes to be validated and performed, while in such cases the fix may simply consist in
replacing some method implementation without introducing any side-effect like changing methods
signatures. Adding cross-cutting concerns, such as logging or security aspects, can also be mostly
useful for capturing, monitoring and managing non-functional behaviours. The ability to dynamically
apply and/or remove aspects in a program is especially appealing, again both in development and
production contexts.

In the case of adaptive execution environments like a Java virtual machine, this however involves
loosing potential optimizations, that can only be obtained and stabilized by a long-running process,
leading to performance deterioration.

\paragraph{Our proposal}
This paper describes a new approach to dynamically patch and weave aspects to a Java
application at run-time by transforming the whole application and making it dynamic.

Our contributions are as follows.

\begin{enumerate}

  \item We designed and developed a Java agent that intercepts and modifies the bytecode at runtime
  to replace  all method invocations by the new Java~7 \texttt{invokedynamic} bytecode.

  \item We propose a control API available through a JMX agent that allows to dynamically manage the
  modifications at runtime without the need for a dedicated language or annotation/pointcut inserts.

  \item We present a comparison between our JooFlux prototype, existing AOP platforms and dynamic
  programming languages. Results demonstrate that JooFlux performances are close to the Java ones
  --- with most of the time a marginal overhead, and sometimes a gain --- while AOP platforms and
  dynamic languages present significant overheads.

\end{enumerate}

\paragraph{Paper structure}
This paper is structured as follows. We start with a contextual overview with brief recalls on
aspect-oriented programming, the Java virtual machine and the new \texttt{invokedynamic}
instruction. Next, we give details on how JooFlux works before comparing its performances with other
platforms and programming languages with dynamic-dispatch. Finally, we discuss the related work and
give perspectives for future work.

\section{Context}

Let us present the context into which this work was conducted. This section contains 3 parts. We
start with an informal recall on aspect-oriented programming. We then continue with an overview of
the Java Virtual Machine. Finally, we focus on a significant evolution made starting from Java SE 7
to ease the support of dynamic languages, and that is at the core of our approach.
\subsection{Aspect-oriented programming}

Aspect-oriented programming (AOP) aims at modularizing the cross-cutting concerns in programs with
most of these being related to non-functional code, e.g. security, validation, logging or transactions
\cite{Kiczales97}. Indeed, such code often needs to be repeated across several application layers,
increasing the maintenance costs as both functional and non-functional code get mixed together.

AOP attempts to solve this problem as follows:
\begin{enumerate}
  \item an \emph{aspect} captures a cross-cutting concern as a single code unit, and
  \item a \emph{point-cut} specifies a set of precise points in a program, and
  \item an \emph{advice} materializes the application of an aspect at a point-cut.
\end{enumerate}

As an example, suppose that we would like to log each invocation to a getter of the classes in
package \texttt{foo}. To do that, an aspect would be the code performing the logging logic, the
point-cut would be the set of calls to public methods of the classes in package \texttt{foo} whose
names start with \texttt{get}. An AOP processor, also called a \emph{code weaver}, would apply the 
aspect to every matching pointcut specified for the advice application.

In practice, the Eclipse AspectJ project is an established AOP framework for Java that features its
own language \cite{Viega2000}. Other approaches exist too \cite{Popovici2002,Chiba2003,Dinn2011}. An
AspectJ source code unit can define aspects, point-cuts and advices. It can also define
\emph{inter-type declarations}, that is, the ability to add new field and method members to existing
types.

Technically, AspectJ works by modifying compiled bytecode either statically by transforming
\texttt{.class} files, or at runtime through the use of a dedicated class loader or JVM agent. In
turn, aspects which are essentially ``just Java code'', get compiled into separate classes. The
AspectJ weaver then applies advices by inserting method calls from the bytecode matching a point-cut
to the corresponding aspect. Back to our example, the logging aspect would be a compiled class, and
each invocation to a matched getter method would be preceded with a invocation of the method holding
the logging logic in the generated aspect.

\subsection{The Java Virtual Machine}

The Java Virtual Machine (JVM) specification has been very stable since its inception
\cite{LindholmJVM99}. A JVM consumes bytecode that is traditionally enclosed into \texttt{.class}
files that represent Java classes. The format of a compiled class was designed to mimic a Java
source file: a compiled \texttt{.class} file contains a single Java class, a set of fields and
methods. It also contains a \emph{constant pool}, that is, a set of indexed constant values that can
be referred to by number in the bytecode, thus reducing the compiled bytecode footprint by avoiding
duplicates.

The execution model of the JVM lies around a stack. Opcodes may manipulate the stack, consume
element as operands, and push new elements. As an example, invoking the \mbox{\texttt{int
java.lang.String::indexOf(String)}} instance method consumes 2 elements from the stack: a
\texttt{java.lang.String} instance to invoke the method followed by another instance that
corresponds to the sole parameter. In return, it pushes a primitive \texttt{int} value as a return
value.

Consider the following Java class:

\begin{Verbatim}[frame=single]
public class Hello {
  public static void main(String... args) {
    System.out.println("Hello!");
  }
}
\end{Verbatim}

Using a decompiler tool such as \texttt{javap}, the bytecode corresponding to the \texttt{main()}
method is as follows:

\begin{Verbatim}[frame=single]
public static void main(java.lang.String...)
  flags: ACC_PUBLIC, ACC_STATIC, ACC_VARARGS
  Code:
    stack=2, locals=1, args_size=1
       0: getstatic     #2                  
       3: ldc           #3                  
       5: invokevirtual #4                  
       8: return        
\end{Verbatim}

The \texttt{getstatic} opcode is used to lookup the \texttt{out} field in the
\texttt{java.lang.System} class which was the second entry from the class constant pool. The
\texttt{ldc} opcode then loads the third value: the \texttt{"Hello"} string. Finally, the
\texttt{println(String)} virtual method is invoked before returning, with the method reference
corresponding to the fourth entry in the constant pool.

While the Java bytecode class format was designed specifically with the requirements of the Java
programming language in mind, it is still a general-purpose bytecode. In fact, the Java bytecode is
strictly more expressive than its counterpart language, a property that has been widely exploited in
applications like obfuscation \cite{Batchelder2007}.

The JVM does not limit itself to running applications written in Java. There is a vibrant ecosystem
of programming languages targeting the JVM. This includes ports of previously-existing languages to
the JVM (e.g., JRuby, Jython, Rhino) or original languages that appeared first on the JVM (e.g.,
Groovy, Scala, Clojure). Indeed, there has been extensive virtual machine optimization research
supported by industrial validation that make the JVM a compelling target runtime for a language
\cite{Paleczny2001,Kotzmann2008,Haubl2011}.

\subsection{Java SE 7 and the new \texttt{invokedynamic} opcode}

The JVM specification provides 4 opcodes for invoking methods. \texttt{invokestatic} is used for
static methods. \texttt{invokevirtual} is used when dispatching shall be performed based on the
receiver type. This corresponds to \texttt{public} and \texttt{protected} methods in Java as they
may be overridden in subclasses. \texttt{invokespecial} is used to dispatch to the other types of
methods such as \texttt{private} methods and constructors. Finally, \texttt{invokeinterface} is used
to dispatch a method to a receiver instance implementing the said interface. This set of method
invocation opcodes remained stable until the release of Java SE 7 when \texttt{invokedynamic} was
introduced \cite{Rose2009}.

\paragraph{A new opcode} The motivation for a new method invocation opcode was to make it easier for
dynamic languages to be implemented on top of the JVM. Indeed, dynamic languages often need to
resolve types, symbols and invocation targets at runtime. Using either of the previous method
invocation opcodes, dynamic language implementors had to rely on reflection and dynamic proxy
generations to delay such tasks to the runtime. This proved to be slow in practice, as just-in-time
optimizations were rarely picked up efficiently by the virtual machines.

\texttt{invokedynamic} is very similar to \texttt{invokeinterface} and \texttt{invokevirtual} in the
sense that the method dispatch is performed at runtime. However, both still require a receiver type,
that is, a base class or interface declaring the target method signature. \texttt{invokedynamic}
relaxes this and is closer to function pointers semantics as, say, in C. More specifically, 
\texttt{invokedynamic} opcodes work with:
\begin{enumerate} 
  \item a symbolic name to designate the invocation, and 
  \item a type signature for the parameters and return type, and 
  \item a bootstrap instruction that is invoked the first time that a given \texttt{invokedynamic}
        opcode is encountered (a call site).
\end{enumerate}

\paragraph{A runtime support API}
The role of the bootstrap instruction is to bind the call site with a target to handle invocations.
To do that, there is a new API found as part of the \texttt{java.lang.invoke} package, and which
defines 2 useful types. \texttt{CallSite} represents a call site, and is the return value of the
bootstrap process. It points to a \texttt{MethodHandle} which is either a direct reference to a
class method or field, or a chain of method handles called \emph{combinators} \cite{Rose2009}. Here
is an usage example of the new API:

\begin{Verbatim}[frame=single]
public static MethodHandle replaceSpaces(Lookup lookup) throws Throwable {
  return insertArguments(lookup.findVirtual(String.class, "replaceAll",
    methodType(String.class, String.class, String.class)), 1, "%20", " ");
}

public static void main(String... args) throws Throwable {
  MethodHandle mh = replaceSpaces(lookup());
  System.out.println((String) mh.invokeExact("A%20B%20C%20"));
}
\end{Verbatim}

It shows how to obtain a method handle over the \texttt{String.replaceAll(String, String)} method.
Once this is done, we insert an \texttt{insertArguments} combinator whose role is to pre-bind
arguments to constant values. In this case we bind both arguments of \texttt{replaceAll()} so that
when invoked, the method replaces occurrences of \texttt{"\%20"} with \texttt{" "}. As this is a
virtual method, the argument at index 0 is actually the receiver object. Finally, the method handle
can be invoked, in this case printing \texttt{"A B C"}. Of course, a wider range of combinators
exists, but the point is that method handles can be manipulated to point to concrete methods or some
transformation / adaptation code as we will see later in this paper.

\begin{figure}[!htb]
  \centering
  \includegraphics[width=\textwidth]{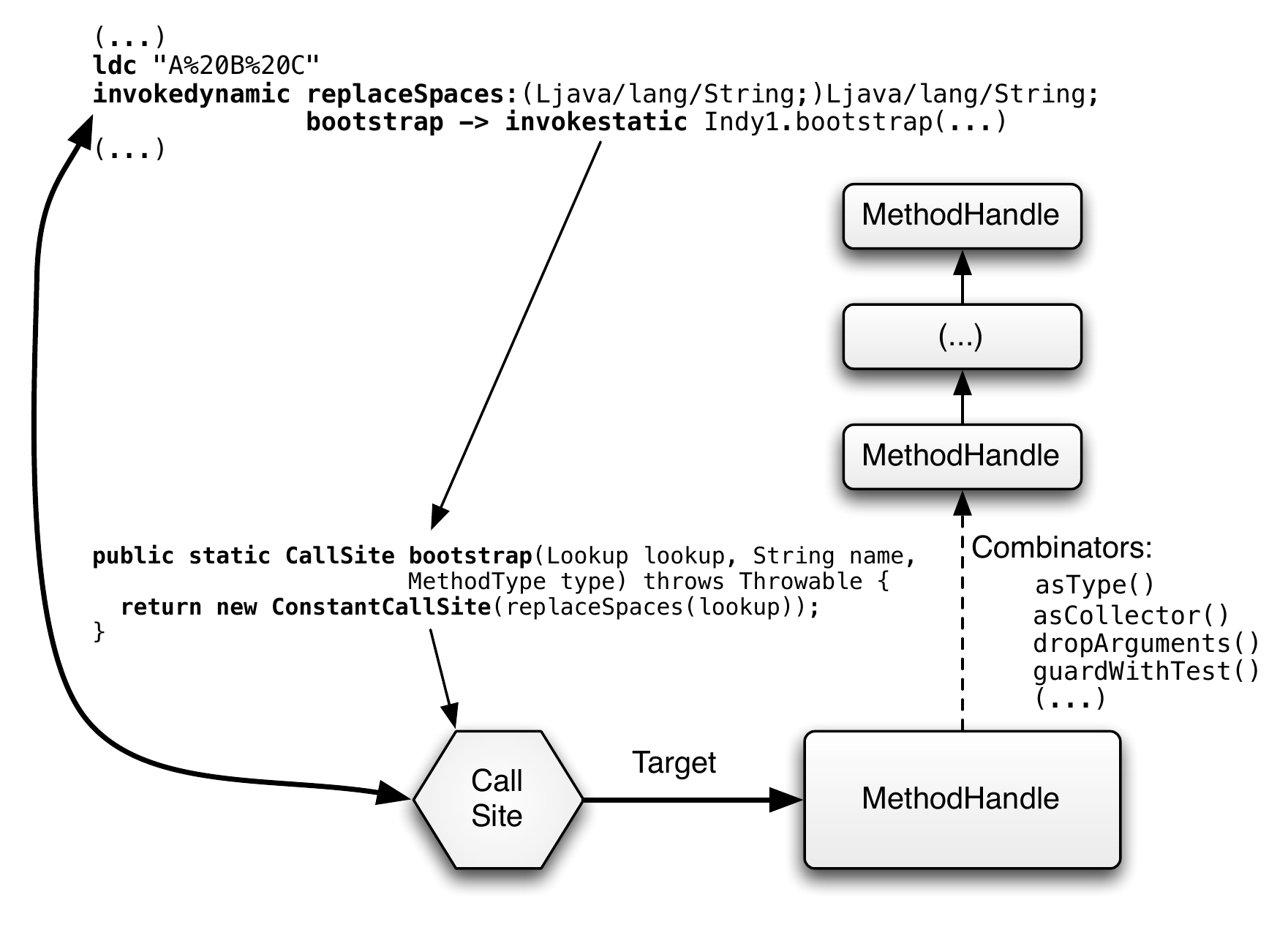}
  \caption{Bootstrapping an invokedynamic call site.}
  \label{fig:indy-bootstrap}
\end{figure}

The link between an \texttt{invokedynamic} opcode and the bootstrap method is
usually done by invoking a static method that obtains a method handle whose type matches those of the call
site, then instantiates a \texttt{CallSite} and returns it. This is illustrated in
Figure~\ref{fig:indy-bootstrap}. The call site instance is only assigned once by invoking the
bootstrap method the first time the invocation is made.

One of the main improvement in using \texttt{invokedynamic} is that combinator chains are assigned
to identifiable call sites and have plugs into the JVM internals. This way, optimizations can be put
in place more easily than with traditional reflective approaches \cite{Thalinger2010}. Another point
is that call sites may be re-targeted to new method handles, which is useful to, say, implement
instance method redefinition at runtime as allowed in languages such as Python or Ruby. Also, method
handles are type-checked only at creation time, unlike reflection objects that need to perform such
checking for each invocation. This yields better method dispatching performance.

\section{JooFlux}

This section gives technical details on how JooFlux works. Specifically, we explain how it
introduces a method call indirection through bytecode rewriting and the \texttt{invokedynamic}
instruction. Then, we explain how aspect advices can be attached to methods, both before and after
they are being executed. Finally, we present the management layer of JooFlux that allows methods to
be replaced, and aspect advices to be injected, all at runtime.

\subsection{Introducing a method call indirection}

JooFlux works by introducing an indirection on method invocations, so that method replacement and
application of aspects can be performed at runtime. While JooFlux primarily focuses on bytecode
emitted by a compiler for the Java programming language, it can theoretically work with any valid
JVM bytecode produced by another language such as Scala~\cite{Odersky05}.

In JooFlux, we took the approach of taking advantage of \texttt{invokedynamic}, as the possibility
of dynamically rebinding call sites to new method handles chains effectively permits method
implementations to be changed. Also, the range of method handle combinators includes what to perform
additional processing on both invocation arguments and return values. As we will see, this
effectively allows us to implement aspect-oriented programming.

As of version 7, Java (the language) does not rely on \texttt{invokedynamic}. Compiled Java bytecode
perform method invocations using the original \texttt{invokestatic}, \texttt{invokevirtual},
\texttt{invokespecial} and \texttt{invokeinterface} opcodes.

\begin{figure}[!htb]
  \centering
  \hspace{-3cm}
  \includegraphics[width=0.85\textwidth]{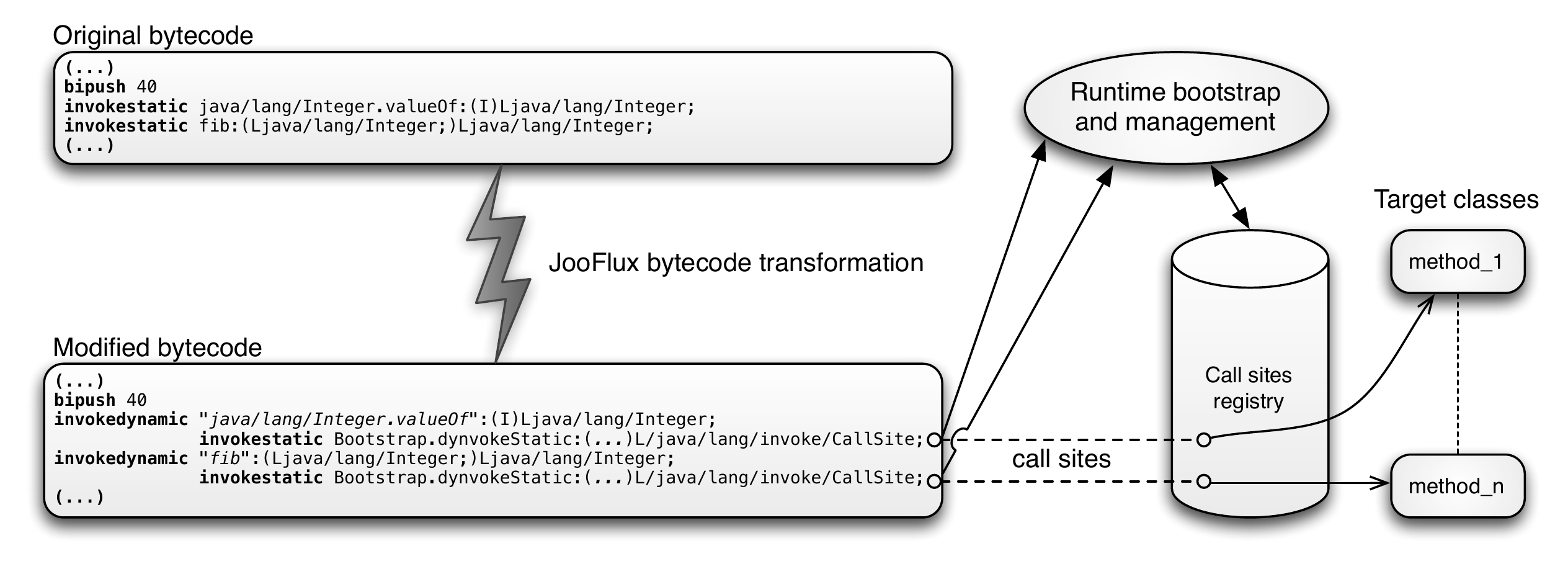}
  \caption{Overview of the JooFlux JVM agent.}
  \label{fig:jooflux-transfo}
\end{figure}

Java virtual machines offer the possibility to attach \emph{agents}. Once plugged into a JVM, an
agent is able to perform many operations, including the ability to define itself as a bytecode
transforming agent. By doing so, an agent can intercept bytecode as it is being loaded into the JVM,
and it can rework it by adding and removing instructions.

JooFlux works as a JVM agent for which Figure~\ref{fig:jooflux-transfo} gives an overview. When
classes are being loaded, it looks for occurrences of \texttt{invokestatic}, \texttt{invokevirtual},
\texttt{invokespecial} or \texttt{invokeinterface} and replaces them by a semantic-preserving
\texttt{invokedynamic} instruction. By doing so, original call sites get bound at runtime to method
handles. The bytecode transformations are being made using the ASM library\footnote{ASM 4.0: 
\url{http://asm.ow2.org/}}.

Initially, the call sites point to method handles to the original classes methods. For instance,
when a \texttt{invokestatic} instruction is in the bytecode, it is replaced by a
\texttt{invokedynamic} instruction whose symbolic name is based on the original
\texttt{invokestatic} target signature.  This is useful to have a uniform naming scheme when we
later want to perform replacements and aspect injection. The type of the original target is
preserved too. The JooFlux runtime call site bootstrap class provides several static methods
depending on the original invocation type, i.e. static, virtual, special and interface. By doing so,
we only introduce a thin indirection layer.

\subsection{Aspect advices using method handle combinators}

In existing tools such as AspectJ, an aspect advice is injected by adding new instructions to the
original bytecode. For instance, an AspectJ advice whose pointcut is before a method invocation will
be applied just before the said method call site. To do so, a first static method is invoked to
fetch the aspect class instance, then a second method invocation is performed on it to the method
that corresponds to the advice. This can be easily checked by decompiling bytecode. Other tools such
as JBoss Byteman work in a similar fashion, while in the case of the later new rules / advices can
be injected and removed at runtime by reloading class definitions. This is allowed by the JVM as
long as the reloaded classes only change method implementations but do not add or remove methods and
fields.

JooFlux does not require adding new instructions into the transformed bytecode. Also, it does not
require reloading classes as new advices are being injected, or when method implementations are
being replaced. Instead, it mainly relies on 2 method handle combinators found in the
\texttt{java.lang.invoke.MethodHandles} class:
\begin{enumerate}
  \item \texttt{filterArguments} takes a target method handle, an argument position and an array of
    filter function method handles, and
  \item \texttt{filterReturnValue} takes a target method handle and a filter function method handle.
\end{enumerate}

The filter method handle function types must match those of their target methods. In our case, we
opted for a generic approach that would fit all kinds of target methods. Hence, JooFlux advices work
with arrays of objects matching the arguments when intercepting invocations, and they use objects
when intercepting returns. The following is an example class providing 2 static methods that can be
used as logging advices:

\begin{Verbatim}[frame=single]
public class Dumpers {
  public static Object[] onCall(Object[] args) {
    System.out.println(">>> " + Arrays.toString(args));
    return args;
  }
  public static Object onReturn(Object retval) {
    System.out.println("<<< " + retval);
    return retval;
  }
}
\end{Verbatim}

In this example the advice methods simply log the invocation arguments and return values. They could
as well do other things such as throwing an exception if some argument value is invalid. Also,
because they act as \emph{filters}, they can modify argument values before they are given to the
target, and they can modify the return value before the invocation client obtains it.

Advices can be stacked, that is, several of them may be applied to a given target method. For
instance, we could stack a validation and a logging advice to be executed before a method is being
invoked. What an advice can do also depends on the target method original invocation mode. In the
case of static methods, the arguments array simply contains the values corresponding to the static
method signature. However in the case of instance methods, the first argument is the receiver, that
is, the instance of the object that the method shall be invoked on.

Finally, we mentioned that target and filter method handles must match. However we opted for a
generic solution that can work with any method signature. To make this work, we took advantage of 2
further method handle combinators. The first one is \texttt{asSpreader}, which can wrap some
arguments into an array. The second one is \texttt{asCollector}, which performs the reverse
operation by mapping the values of an array to parameters. For method invocation interception, we
first wrap all arguments into a \texttt{Object[]} array using \texttt{asSpreader}. We can then pass
it to the advice method, and extract the filtered values from the resulting array back to the target
method parameters using \texttt{asCollector}. The case of method return interception is simpler as
it does not require using such combinator chain. Instead, we just need to loosen the return value
type to \texttt{Object} before passing it to the advice, then narrow it again to its original type
before returning to the method invoker. Such type transformations happen with the \texttt{asType}
combinator.

\subsection{Managing JooFlux: live method replacement and aspects injection}

Call sites are being put in a central registry at bootstrap time. This registry sports no role while
application code is being executed, and there is also no lookup penalty in having it. It is used by
the management layer of JooFlux in several ways. To do so, JooFlux offers a JMX\footnote{Java Management Extensions: \url{http://jcp.org/aboutJava/communityprocess/final/jsr003/index3.html}} agent for querying and
interacting with itself.

This agent offers the following set of remotely-accessible operations:
\begin{enumerate}
  \item replacing method implementations by changing their call site targets, and
  \item plugging an aspect advice before or after certain call sites, and
  \item querying for various metrics, including the number of call sites monitored by JooFlux and a
    list of them.
\end{enumerate}

As an example, let us see the interaction with the method replacement operation, whose signature as
a JMX interface method is 

\begin{Verbatim}[frame=single]
void changeCallSiteTarget(String methodType, 
                          String oldTarget, String newTarget)
\end{Verbatim}

The \texttt{methodType} parameter specifies the type of method invocation in the
original bytecode: static, virtual, interface and special. The last 2 parameters specify an
identifier for a call site to be replaced and a new method handle to use as a target.

\begin{figure}[!htb]
  \centering
  \includegraphics[width=\textwidth]{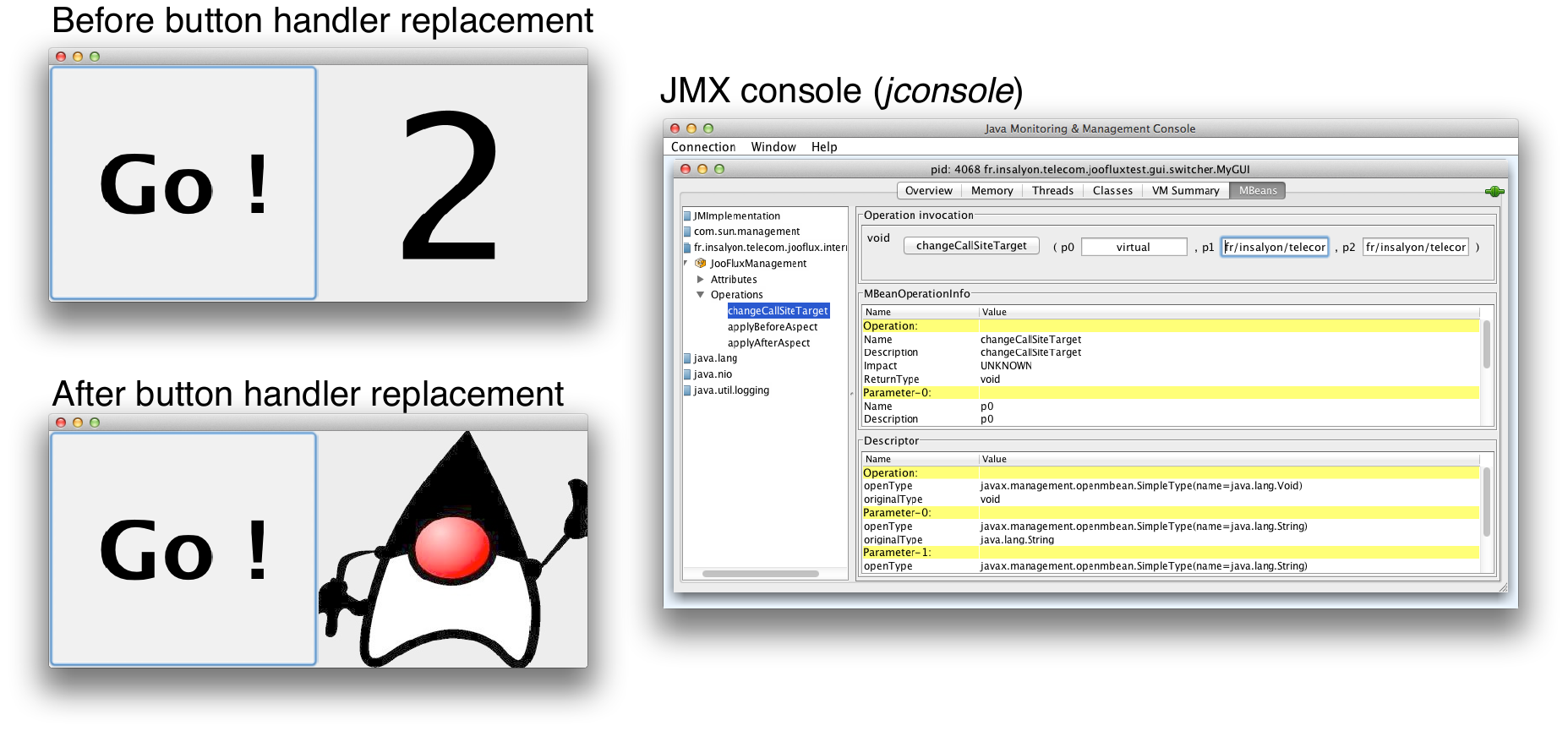}
  \caption{Interacting with an application through the JooFlux JMX agent.}
  \label{fig:mbean}
\end{figure}

Figure~\ref{fig:mbean} features a graphical application with a button on the left pane and a text
label on the right pane. By connecting to the JooFlux JMX agent of the application using a tool such
as \emph{jconsole}, we can replace the button action handler so that the next time it is clicked, a
picture replaces the label. In that specific demonstration case, we used the following input to the
\texttt{changeCallSiteTarget} JMX operation: 

\begin{Verbatim}[frame=single]
virtual, // virtual method
fr/insalyon/telecom/joofluxtest/gui/switcher/
   MyActionListener.counterIncrement:(MyActionListener)void, // handler id
fr/insalyon/telecom/joofluxtest/gui/switcher/
   MyActionListener.pictureSwitch:()V // our new handler
\end{Verbatim}

Injecting an aspect into an existing call site works in a similar fashion:

\begin{Verbatim}[frame=single]
void applyBeforeAspect(String callSitesKey,
                       String aspectClass, String aspectMethod)

void applyAfterAspect(String callSitesKey, 
                      String aspectClass, String aspectMethod)
\end{Verbatim}


\section{Experimental comparison with AOP platforms and dynamic languages}
\label{section:bench}

To assess our approach, we compare the performance of the Java Virtual Machine with our JooFlux agent to other dynamic method call dispatching approaches: AOP platforms~(section~\ref{section:aop}) and dynamic programming languages~(sections~\ref{section:microdynlang}-\ref{section:macrodynlang}). 

All these approaches are JVM-based and have been tested with micro- and macro-benchmarks on a MacBook~Pro~2,3~GHz~Intel~Core2~(i5), 4~Go~1333~MHz~DDR3, running Mac~OS~X~Lion~10.7.4~(11E53) and the OpenJDK Runtime Environment~1.7.0-u10-b06-20120906\footnote{OpenJDK Runtime Environment~1.7.0-u10-b06-20120906: \url{http://code.google.com/p/openjdk-osx-build/}}. These tests were renewed 10~times to constitute the result set. Quartiles and median overhead were calculated on this result set.

\subsection{Micro-benchmarks and AOP platforms}
\label{section:aop}

We compare JooFlux aspect injection functionality with two other AOP platforms: \mbox{AspectJ}\footnote{AspectJ~1.7.1: \url{http://www.eclipse.org/aspectj/}}\cite{KiczalesHHKPG01} and \mbox{Byteman}\footnote{Byteman~2.1.0: \url{http://www.jboss.org/byteman/}}\cite{Dinn2011}. We use a classic recursive Fibonacci micro-benchmark (\texttt{classicfibo})\footnote{Recursive Fibonacci: \url{http://en.wikipedia.org/wiki/Recursion\_(computer\_science)\#Fibonacci}} where we inject an empty aspect - i.e. a redirection call to an empty method - before or/and after the Fibonacci method. Results are presented in Table~\ref{tab:fiboaop}.

\begin{table}[!htb]
\centering
\begin{tabular}{|c|c|c|c|c|c|c|c||c||}
\hline
\hline
Exec. Plat. & Weav. Time & Impl. & Q1\footnotesize{-min} & Q2\footnotesize{-25\%} & Q3\footnotesize{-median} & Q4\footnotesize{-75\%} & Q5\footnotesize{-max} & Overhead \\
\hline
\hline
JVM & - & classicfibo & 611 & 613 & 615 & 616 & 636 & - \\
\hline
& \multirow{7}{*}{compilation} & classicfibo & \multirow{2}{*}{609} & \multirow{2}{*}{613} & \multirow{2}{*}{620} & \multirow{2}{*}{621} & \multirow{2}{*}{650} & \multirow{2}{*}{+0,8\%} \\ 
& & + before aspect & & & & & & \\
\cline{3-9}
\multirow{2}{*}{JVM +} & & classicfibo & \multirow{2}{*}{609} & \multirow{2}{*}{611} & \multirow{2}{*}{620} & \multirow{2}{*}{621} & \multirow{2}{*}{693} & \multirow{2}{*}{+0,8\%} \\
\multirow{2}{*}{AspectJ} & & + after aspect & & & & & & \\
\cline{3-9}
& & classicfibo & \multirow{3}{*}{609} & \multirow{3}{*}{613} & \multirow{3}{*}{620} & \multirow{3}{*}{622} & \multirow{3}{*}{717} & \multirow{3}{*}{+0,8\%} \\
& & + before aspect & & & & & & \\
& & \& after aspect & & & & & & \\
\hline
& \multirow{7}{*}{runtime} & classicfibo & \multirow{2}{*}{3007} & \multirow{2}{*}{3028} & \multirow{2}{*}{3051} & \multirow{2}{*}{3070} & \multirow{2}{*}{3525} & \multirow{2}{*}{+396\%} \\ 
& & + before aspect & & & & & & \\
\cline{3-9}
\multirow{2}{*}{JVM +} & & classicfibo & \multirow{2}{*}{1269} & \multirow{2}{*}{1273} & \multirow{2}{*}{1280} & \multirow{2}{*}{1295} & \multirow{2}{*}{1356} & \multirow{2}{*}{+108\%} \\
\multirow{2}{*}{JooFlux Agent} & & + after aspect & & & & & & \\
\cline{3-9}
& & classicfibo & \multirow{3}{*}{3804} & \multirow{3}{*}{3830} & \multirow{3}{*}{3861} & \multirow{3}{*}{3893} & \multirow{3}{*}{3950} & \multirow{3}{*}{+528\%} \\
& & + before aspect & & & & & & \\
& & \& after aspect & & & & & & \\
\hline
& \multirow{7}{*}{runtime} & classicfibo & \multirow{2}{*}{79309} & \multirow{2}{*}{84998} & \multirow{2}{*}{85371} & \multirow{2}{*}{85695} & \multirow{2}{*}{86397} & \multirow{2}{*}{+13781\%} \\ 
& & + before aspect & & & & & & \\
\cline{3-9}
\multirow{2}{*}{JVM +} & & classicfibo & \multirow{2}{*}{82561} & \multirow{2}{*}{82780} & \multirow{2}{*}{82911} & \multirow{2}{*}{83040} & \multirow{2}{*}{83406} & \multirow{2}{*}{+13381\%} \\
\multirow{2}{*}{Byteman} & & + after aspect & & & & & & \\
\cline{3-9}
& & classicfibo & \multirow{3}{*}{179984} & \multirow{3}{*}{188088} & \multirow{3}{*}{188187} & \multirow{3}{*}{188321} & \multirow{3}{*}{188971} & \multirow{3}{*}{+30499\%} \\
& & + before aspect & & & & & & \\
& & \& after aspect & & & & & & \\
\hline
\hline
\end{tabular}
\caption{Fibo(40) micro-benchmarks for AOP platforms (in ms)}
\label{tab:fiboaop}
\end{table}

Execution times of the \mbox{AspectJ} generated bytecodes are close to the JVM ones, but it weaves the aspects statically during the compilation and no modification can be performed at runtime. \mbox{Byteman} allows the aspect injection at runtime by unloading the class, and modifies the bytecode during the reloading. This technique degrades significantly the performance and implies that all optimizations achieved by the JIT are lost. The bytecode generated by \mbox{JooFlux} already includes dynamic calls, so the cost of weaving an aspect is just the cost of including a new combinator in the method handle chains, recopying the arguments and transferring the return value. By keeping the method handle chain almost intact, \mbox{JooFlux} protects the JIT optimizations.

\subsection{Micro-benchmarks and dynamic languages}
\label{section:microdynlang}

Several programming languages propose the dynamic method dispatch directly in the language, e.g. with reflective APIs. To test the JooFlux dynamic method dispatch functionality, we compare it to the main JVM-based programming languages: Java\footnote{Java~7: \url{http://docs.oracle.com/javase/7/docs/api/}}, Clojure\footnote{Clojure~1.4.0: \url{http://clojure.org}}, JRuby\footnote{JRuby~1.6.7.2~(-indy), 1.7.0.preview2~(+indy): \url{http://jruby.org}}, Groovy\footnote{Groovy~2.0.2~(-+indy): \url{http://groovy.codehaus.org}}, Rhino~JavaScript\footnote{Rhino~Javascript~1.7R4: \url{https://developer.mozilla.org/en-US/docs/Rhino}} and Jython\footnote{Jython~2.5.3: \url{http://www.jython.org}}. \\

We first use the same Fibonacci micro-benchmark to test the performance. According to the
programming languages,  the Fibonacci method can be written in different manner: the
\texttt{classicfibo} that manipulates objects, the \texttt{fastfibo} that manipulates long-typed
parameters, the \texttt{fastestfibo} that manipulates long-typed parameters and return long-typed
results and finally the \texttt{reflectivefibo} where each method invocation use the reflective~API.
These different implementations are important because they influence the generated bytecode and
hence performance.

Java results are presented in Table~\ref{tab:fibojava}. JooFlux overhead is insignificant for \texttt{classicfibo} and presents a slowing factor of~2 for \texttt{reflectivefibo}. This implementation is indeed our worst case as the JVM can hardly inline the method calls.


\begin{table}[!htb]
\centering
\begin{tabular}{|c|c|c|c|c|c|c|c||c||}
\hline
\hline
P. Lang. & Exec. Plat. & Impl. & Q1\footnotesize{-min} & Q2\footnotesize{-25\%} & Q3\footnotesize{-median} & Q4\footnotesize{-75\%} & Q5\footnotesize{-max} & Overhead \\
\hline
\hline
\multirow{2}{*}{Java} & \multirow{2}{*}{JVM} & classicfibo & 611 & 613 & 615 & 616 & 636 & - \\
\cline{3-9}
& & reflectivefibo & 1758 & 1762 & 1782 & 1803 & 4121 & - \\
\hline
\multirow{2}{*}{Java} & JVM + & classicfibo & 611 & 613 & 616 & 618 & 690 & +0,001\% \\ 
\cline{3-9}
& JooFlux Agent & reflectivefibo & 3668 & 3686 & 3717 & 3743 & 4273 & +108\% \\
\hline
\hline	
\end{tabular}
\caption{Fibo(40) micro-benchmarks for Java programming language (in ms)}
\label{tab:fibojava}
\end{table}

The others programming languages results are presented in Table~\ref{tab:fibodynlang}. The different \texttt{classicfibo}, \texttt{fastfibo} and \texttt{fastestfibo} overheads are calculated compared to JVM+JooFlux \texttt{classicfibo}. \texttt{reflectivefibo} implementations are still compared to JVM+JooFlux \texttt{reflectivefibo} one.

Even if Clojure strongly-typed implementations presents only a slowdown factor of 1.2-1.4, most other languages are from 3~to 18~times slower than our Java+JooFlux prototype. JRuby and Groovy languages propose earlier versions using the \texttt{invokedynamic} bytecode \textit{(+indy, figures in italic in the Table~\ref{tab:fibodynlang})} but even if they gain a slowing factor, they remain significantly slower than Java+JooFlux.

\begin{table}[!htb]
\centering
\begin{tabular}{|c|c|c|c|c|c|c|c||c||}
\hline
\hline
P. Lang. & Exec. Plat. & Impl. & Q1\footnotesize{-min} & Q2\footnotesize{-25\%} & Q3\footnotesize{-median} & Q4\footnotesize{-75\%} & Q5\footnotesize{-max} & JooFlux Diff \\
\hline
\hline
\multirow{3}{*}{Clojure} & \multirow{3}{*}{JVM} & fastestfibo & 722 & 732 & 734 & 740 & 742 & +19\% \\
\cline{3-9}
& & fastfibo & 859 & 862 & 864 & 875 & 892 & +40\% \\
\cline{3-9}
& & classicfibo & 4105 & 4118 & 4171 & 4265 & 4326 & +577\% \\
\hline
\multirow{4}{*}{JRuby} & \multirow{3}{*}{JVM} & \multirow{2}{*}{classicfibo} & 6290 & 6333 & 6382 & 6486 & 7043 & +936\% \\
& & & \it{(3982)} & \it{(4006)} & \it{(4020)} & \it{(4069)} & \it{(4323)} & \it{(+552\%)} \\
\cline{3-9}
& \it{(+indy)} & \multirow{2}{*}{reflectivefibo} & 10226 & 12020 & 12060 & 12076 & 12288 & +224\% \\
& & & \it{(7545)} & \it{(7561)} & \it{(7581)} & \it{(7621)} & \it{(7737)} & \it{(+104\%)} \\
\hline
\multirow{8}{*}{Groovy} & \multirow{8}{*}{JVM} & \multirow{2}{*}{fastestfibo} & 1383 & 1388 & 1394 & 1401 & 1417 & +126\% \\
& & & \it{(3061)} & \it{(3077)} & \it{(3092)} & \it{(3150)} & \it{(3165)} & \it{(+402\%)} \\
\cline{3-9}
& & \multirow{2}{*}{fastfibo} & 2709 & 2721 & 2725 & 2749 & 2766 & +342\% \\
& & & \it{(2513)} & \it{(2519)} & \it{(2528)} & \it{(2540)} & \it{(2583)} & \it{(+310\%)} \\
\cline{3-9}
& \multirow{2}{*}{\it{(+indy)}} & \multirow{2}{*}{classicfibo} & 8660 & 8691 & 8716 & 8726 & 9066 & +1315\% \\
& & & \it{(4461} & \it{(4488)} & \it{(4522)} & \it{(4584)} & \it{(4656)} & \it{(+634\%)} \\
\cline{3-9}
& & \multirow{2}{*}{reflexivefibo} & 57734 & 57892 & 58009 & 58182 & 58364 & +1460\% \\
& & & \it{(8366} & \it{(8378)} & \it{(8386)} & \it{(8405)} & \it{(8697)} & \it{(+125\%)} \\
\hline
Javascript & JVM & classicfibo & 9052 & 9208 & 11275 & 11441 & 11764 & +1730\% \\
\hline
Jython & JVM & classicfibo & 29053 & 29258 & 29675 & 30202 & 31871 & +4717\% \\
\hline
\hline	
\end{tabular}
\caption{Fibo(40) micro-benchmarks for JVM-based dynamic languages (in ms)}
\label{tab:fibodynlang}
\end{table}

For AOP and dynamic languages Fibonacci micro-benchmarks, our JooFlux agent rewrites the bytecode in
75-100ms for a transformation of 2~classes~(\texttt{Fibonacci} and \texttt{InvokeBootstrap} - the
bootstrap itself is dynamic) and 1~method~(\texttt{classicfibo}). This delay is introduced before
the application launch, so it is quite insignificant and comparable to other platforms launching times.

\subsection{CPU-intensive and interception/rewrite-intensive macro-benchmarks}
\label{section:macrodynlang}

To test more intensively the JooFlux dynamic method dispatch, we have used 3~macro-benchmarks:
\begin{enumerate}
\item SCImark 2.0\footnote{SCImark 2.0: \url{http://math.nist.gov/scimark2/}}: SciMark 2.0 is a CPU- and memory-intensive Java benchmark for scientific and numerical computing. It measures several computational kernels and reports a composite score in approximate Mflops.
\item A parallel file wordcounter with Fork/Join~\cite{ponge:inria-00611456}: This file wordcounter is a memory- and IO-intensive Java benchmark that looks down a repository hierarchy and counts words inside each file.
\item The execution of Clojure runtime over JooFlux: We have used the Clojure language runtime -- written in Java -- to have an Interception-intensive benchmark.
\end{enumerate}

\paragraph{SCImark 2.0}

The benchmark performs Fast Fourier Transformations~(FFT), Jacobi Successive Over-relaxation~(SOR), Monte Carlo integration~(MC), Sparse matrix multiply~(SM), dense LU matrix factorization~(LU) and computes a composite score~(CS). These calculations can be done with a small- or large-memory data set.

Performance results with and without-JooFlux are presented in Table~\ref{tab:scimark2java}. Almost all calculations have a positive or negative marginal overhead, except the Monte Carlo integration that loses 20\% performance. The Monte Carlo algorithm exercises random-number generators, synchronized function calls, and function inlining that can explain the JIT difficulties.

\begin{table}[!htb]
\centering
\begin{tabular}{|c|c|c|c|c|c|c|c|c|c|c|c|}
\hline
\hline
& \multicolumn{2}{c|}{FFT} & \multicolumn{2}{c|}{SOR} & MC & \multicolumn{2}{c|}{SM} & \multicolumn{2}{c|}{LU} & \multicolumn{2}{c|}{CS} \\
\cline{2-12}
& \multirow{2}{*}{$\!\!\!2^{10}\!\!\!$} & \multirow{2}{*}{$\!\!\!2^{20}\!\!\!$} & \multirow{2}{*}{$\!\!10^{2}\!\!\times\!\!10^{2}\!\!$} & \multirow{2}{*}{$\!\!10^{3}\!\!\times\!\!10^{3}\!\!$} & \multirow{2}{*}{-} & $\!\!\!N\!\!=\!\!10^{3}\!\!\!$ & $\!\!\!N\!\!=\!\!10^{5}\!\!\!$ & \multirow{2}{*}{$\!\!10^{2}\!\!\times\!\!10^{2}\!\!$} & \multirow{2}{*}{$\!\!10^{3}\!\!\times\!\!10^{3}\!\!$} & \multirow{2}{*}{Small} & \multirow{2}{*}{Large} \\
& & & & & & $\!\!nz\!\!=\!\!5.10^{3}\!\!\!$ & $\!\!\!nz\!\!=\!\!10^{6}\!\!\!$ & & & & \\
\hline
JVM & 827 & 178,7 & 1196,5 & 1077,6 & 635,4 & 1180,8 & 1227,8 & 2467,8 & 1490,2 & 1261,5 & 915,4 \\
\hline
JVM + & \multirow{2}{*}{824,3} & \multirow{2}{*}{178,5} & \multirow{2}{*}{1193,2} & \multirow{2}{*}{1080,3} & \multirow{2}{*}{507,2} & \multirow{2}{*}{1172,9} & \multirow{2}{*}{1249,5} & \multirow{2}{*}{2325,9} & \multirow{2}{*}{1489,7} & \multirow{2}{*}{1204,9} & \multirow{2}{*}{900,2} \\
JooFlux & & & & & & & & & & & \\
\hline
\hline	
Perf. & -0,3\% & -0,1\% & -0,3\% & +0,3\% & -20\% & -0,7\% & +1,8\% & -5,8\% & -0,03\% & -4,5\% & -1,7\% \\
\hline
\hline
\end{tabular}
\caption{SCImark2.0 small and large macro-benchmarks (in Mflops)}
\label{tab:scimark2java}
\end{table}

\paragraph{Parallel file wordcounter with Fork/Join}

We apply the file wordcounter to the HotSpotVM source repository~($2.10^{3}$~files) with a mono-thread process and with a 2-threads on 2-cores parallel execution. As shown in Table~\ref{tab:forkjoinjava}, surprisingly, the bytecode modified by JooFlux betters the performances. This result demonstrates that synchronized method calls are not affected by the \texttt{invokedynamic} indirection and the JIT can even apply a better inlining.

\begin{table}[!htb]
\centering
\begin{tabular}{|c|c|c|c|c|c|c||c||}
\hline
\hline
Exec. Plat. & Impl. & Q1\footnotesize{-min} & Q2\footnotesize{-25\%} & Q3\footnotesize{-median} & Q4\footnotesize{-75\%} & Q5\footnotesize{-max} & Overhead \\
\hline
\hline
\multirow{4}{*}{JVM} & wordcounter & \multirow{2}{*}{3707} & \multirow{2}{*}{3727} & \multirow{2}{*}{3733} & \multirow{2}{*}{3752} & \multirow{2}{*}{4769} & \multirow{2}{*}{-} \\
& single thread & & & & & & \\
\cline{2-8}
& wordcounter & \multirow{2}{*}{1826} & \multirow{2}{*}{1934} & \multirow{2}{*}{2161} & \multirow{2}{*}{2235} & \multirow{2}{*}{5949} & \multirow{2}{*}{-} \\
& fork/join 2 threads & & & & & & \\
\hline
& wordcounter & \multirow{2}{*}{3634} & \multirow{2}{*}{3646} & \multirow{2}{*}{3658} & \multirow{2}{*}{3689} & \multirow{2}{*}{4829} & \multirow{2}{*}{-2\%} \\ 
JVM + & single thread & & & & & & \\
\cline{2-8}
JooFlux Agent & wordcounter & \multirow{2}{*}{1825} & \multirow{2}{*}{1916} & \multirow{2}{*}{2029} & \multirow{2}{*}{2070} & \multirow{2}{*}{2407} & \multirow{2}{*}{-6\%} \\
& fork/join 2 threads & & & & & & \\
\hline
\hline	
\end{tabular}
\caption{Parallel file wordcounter macro-benchmarks for Java programming language (in ms)}
\label{tab:forkjoinjava}
\end{table}

\paragraph{Clojure over JooFlux}

The Clojure language runtime is a quite large Java application~($3.10^{3}$~classes) and we use it as an interception-intensive benchmark. We have retested the Fibonacci micro-benchmark, but this time with a total Clojure rewritten bytecode: 1325 transformed class, 26866 transformed methods and 19646 initial interceptions. We do not log the total number of dynamic method call interceptions as continuous loggers dramatically slow down the performance.

Table~\ref{tab:fiboclojure}\footnote{Fibo(40) in Clojure does not have exactly the same performance in Table~\ref{tab:fiboclojure} and Table~\ref{tab:fibodynlang}. To allow a good comparison, we prefer presenting Clojure+JVM results tested at the same time as Clojure+JVM+JooFlux, in the same test run.} presents the results. The overhead introduces is still insignificant compared to the execution time. We can say so that having a dynamic Clojure is effectively nice\footnote{Why Clojure doesn’t need invokedynamic, but it might be nice: \\ \url{http://blog.fogus.me/2011/10/14/why-clojure-doesnt-need-invokedynamic-but-it-might-be-nice/}} and has no cost at runtime ! The 4s bytecode transformation time is not any more negligible but this cost is only applied one time at launch time and it's a low price to let the entire application becomes dynamic !

\begin{table}[!htb]
\centering
\begin{tabular}{|c|c|c|c|c|c|c|c||c||}
\hline
\hline
P. Lang. & Exec. Plat. & Impl. & Q1\footnotesize{-min} & Q2\footnotesize{-25\%} & Q3\footnotesize{-median} & Q4\footnotesize{-75\%} & Q5\footnotesize{-max} & Overhead \\
\hline
\hline
\multirow{3}{*}{Clojure} & \multirow{3}{*}{JVM} & fastestfibo & 701 & 702 & 704 & 708 & 717 & - \\
\cline{3-9}
& & fastfibo & 839 & 845 & 848 & 851 & 855 & - \\
\cline{3-9}
& & classicfibo & 3969 & 3978 & 3984 & 4010 & 4093 & - \\
\hline
\multirow{3}{*}{Clojure} & \multirow{2}{*}{JVM +} & fastestfibo & 695 & 699 & 702 & 704 & 713 & -0,3\% \\
\cline{3-9}
& \multirow{2}{*}{JooFlux Agent} & fastfibo & 856 & 861 & 864 & 868 & 891 & +1,9\% \\
\cline{3-9}
& & classicfibo & 3957 & 3969 & 3980 & 4002 & 4069 & -0,1\% \\
\hline
\hline	
\end{tabular}
\caption{Fibo(40) for Clojure over a JVM+JooFlux execution plate-form (in ms)}
\label{tab:fiboclojure}
\end{table}

\section{Related work}

\paragraph{Software dynamic updates} The ability to dynamically update software at runtime is
anything but a new idea, and it has been investigated by many research works
\cite{Frieder1991,Hicks2005}. In fact, the techniques to be leveraged vastly dependent on target the
programming language semantics and runtime. Dynamic software updates has been investigated as low as
the operating system kernel level in works such as \cite{Arnold2009} and \cite{Palix2011}. Closer to
the context of applications running on a Java virtual machines, the work in \cite{Wurthinger2010}
studies applying modifications of Java code at runtime. The approach supports modification of
classes, including adding and removing fields and methods. However, this requires the usage of a
modified virtual machine. The same authors continued their approach based on virtual machine
modifications in \cite{Wurthinger2011}, and applied it to dynamic aspect-oriented programming. It
should also be mentioned that class definitions may be reloaded in the JVM. This is what debuggers
do, and JVM agents such as the one provided by JBoss Byteman leverages this mechanism. This works
only with stable class definitions though, and accumulated JIT optimizations get lost on reload.

\paragraph{Aspect-oriented programming}
Many works focus on the design and implementation of aspect-oriented programming. AspectJ is a
well-known tool comprising a compiler and language \cite{KiczalesHHKPG01}. While static, it allows
for efficient bytecode transformation when weaving aspects, and offers a fine-grained language to
define pointcuts. Javassist is another proposal to do AOP in Java \cite{Chiba2003}. Some works focus
on the ability to weave aspects in a dynamic fashion \cite{Popovici2002,Popovici2003}. JBoss Byteman is
particularly interesting as it features a \emph{event-condition-action} rule language
\cite{Dinn2011}. Also, it can attach its JVM agent to already running applications as well as apply
or remove rules dynamically. The work in \cite{Bockisch2004} further illustrates that providing
modified virtual machines is a sound solution to support a complete range of modification operations
to running Java applications. This concords with the results in \cite{Wurthinger2011}.

\paragraph{Virtual machines} The field of virtual machines is no short of challenges. The JVM is an
attractive platform for research given its open specification \cite{LindholmJVM99}. Many works focus
on improving the JVMs performance \cite{Paleczny2001,Kotzmann2008,Haubl2011}. The design of the JVM
has an initial bias towards favouring statically typed languages. The rise of interest in dynamic
languages running on the JVM yield to the \texttt{invokedynamic} instruction and the
\texttt{java.lang.invoke} API \cite{Rose2009}.  This facilitates the implementation of such
languages, and it also gives efficient plugs into the JVM internals so as to benefit from adaptive
optimizations \cite{Thalinger2010}. New usages of \texttt{invokedynamic} are starting to appear
\cite{Appeltauer2010}, and it is poised to serve as an implementation technique for the support of
\emph{lambdas} in Java 8\footnote{See \url{http://openjdk.java.net/projects/lambda/}}.

\section{Conclusions and perspectives}

We conclude this paper with some perspectives envisioned at this early stage of the project, and
give pointers to obtain a copy of our prototype.

\subsection{Conclusion}

This paper introduced JooFlux, a JVM agent that allows both the dynamic replacement of method
implementations and the application of aspect advices. Compared to existing approaches, JooFlux
takes a novel route by taking advantage of the new \texttt{invokedynamic} instruction added in Java
SE 7. To do so, it performs bytecode rewriting by introducing an \texttt{invokedynamic}-based
indirection on method invocations, which is later used to dynamically modify the intercepted call
sites. As shown in the earlier micro-benchmarks, the runtime overhead of JooFlux is marginal for
method invocations, and fairly limited when aspects are being injected. In any case, JooFlux shows
interesting performance compared to related approaches such as AOP tools or dynamic languages that
rely on dynamic dispatch. More interestingly, JooFlux does not involve reloading whole classes on
either method replacement or advice injection, which keeps a large range of just-in-time compilation
optimizations valid. It does not require a specifically-tailored virtual machine. Also, the thin
indirection layer introduced by JooFlux does not require lookups or guard checks for call site
invalidation, greatly helping virtual machines in adaptive optimization work. Finally, JooFlux works
in a transparent fashion by directly operating at the method call site level. As such, it does not
require a dedicated language to specify where and what changes to apply.

\subsection{Future work}

JooFlux is currently a research prototype that demonstrates how \texttt{invokedynamic} can be
cleverly used for other purposes than implementing dynamic languages on the JVM. While it
demonstrates and validates our approach with small testing code bases, it still needs to be tested
on a wider range of applications running on the JVM. The impact of the bytecode manipulation that we
perform is significant and we are well aware that it can break some bytecode constructions. As more
testing is being made, we hope to iron out the large share of possible corner cases. By doing so, we
intend to turn JooFlux into a tool that can be adopted at a larger scale than what our initial
version allows.

Given the promising performance figures exhibited by JooFlux, we intend to study its application in
specialized contexts such as resource control, multi-tenant architectures or dynamic modular
applications for the \emph{Internet of Things} without the need to rely on dedicated middleware
platforms such as OSGi\footnote{See \url{http://www.osgi.org/}}.

In terms of general purpose features, we intend to introduce modification transactions and rollback.
The ability to statically verify the applicability of modifications is appealing. Also, we need to
improve the tooling to make it more convenient than a JMX agent. A secure remote shell interface
through SSH would be especially convenient, as it would also support new code remote transport over
tunnelling.

\subsection{Availability}

JooFlux is available as an open source project at \url{https://github.com/dynamid/jooflux}. It is
published under the terms of the \emph{Mozilla Public License Version 2.0}\footnote{See
\url{http://www.mozilla.org/MPL/2.0/}}. The prototype version of JooFlux used while writing this
paper corresponds to the \texttt{r0} annotated tag of the corresponding Git repository. The source
code contains build and testing instructions, and it also contains some instructions for reproducing
demonstrations such as the one related in Figure~\ref{fig:mbean}. It also contains the scripts or
instructions to reproduce the experiments made in Section~\ref{section:bench}.

As further developments are being made, the range of features described in this paper may have
changed, possibly also impacting performance figures. We encourage the wider researchers and
practitioners community to report any issue with JooFlux, and contribute bug fixes and improvements.

\bibliographystyle{plain}
\bibliography{biblio}

\end{document}